\documentclass[%
reprint,
superscriptaddress,
 amsmath,amssymb,
 aps,
prl,
]{revtex4-2}

\usepackage{graphicx}
\usepackage{dcolumn}
\usepackage{bm}
\usepackage{xcolor}
\usepackage[normalem]{ulem}

\setcitestyle{super, open={},close={}}

\begin{document}
\preprint{APS/123-QED}

\title{The pseudochiral Fermi surface of $\alpha$-RuI$_3$}

\author{Alex Louat}
\email{alex.louat@diamond.ac.uk}
\affiliation{Diamond Light Source, Harwell Science and Innovation Campus, Didcot OX11 0DE, UK}%
\author{Matthew D. Watson}%
\affiliation{Diamond Light Source, Harwell Science and Innovation Campus, Didcot OX11 0DE, UK}%
\author{Timur K. Kim}
\affiliation{Diamond Light Source, Harwell Science and Innovation Campus, Didcot OX11 0DE, UK}%
\author{Danrui Ni}
\affiliation{Department of Chemistry, Princeton University\\Princeton, NJ 08544, USA}%
\author{Robert J. Cava}
\affiliation{Department of Chemistry, Princeton University\\Princeton, NJ 08544, USA}%
\author{Cephise Cacho}
\email{cephise.cacho@diamond.ac.uk}
\affiliation{Diamond Light Source, Harwell Science and Innovation Campus, Didcot OX11 0DE, UK}%

\date{\today}


\begin{abstract}
\begin{center}\textbf{Abstract}\end{center}

In continuation of research into RuCl$_3$ and RuBr$_3$ as potential quantum spin liquids, a phase with unique magnetic order characterised by long-range quantum entanglement and fractionalised excitations, the compound RuI$_3$ has been recently synthesised. Here, we show RuI$_3$ is a moderately correlated metal with two bands crossing the Fermi level, implying the absence of any quantum spin liquids phase. We find that the Fermi surface as measured or calculated for a 2D ($k_\text{x},k_\text{y}$) slice at any $k_\text{z}$ lacks mirror symmetry, i.e. is pseudochiral. We link this phenomenon to the ABC stacking in the R$\bar{3}$ space group of $\alpha$-RuI$_3$, which is achiral but lacks any mirror or glide symmetries. We further provide a formal framework for understanding when such a pseudochiral electronic structure may be observed.

\end{abstract}

\maketitle

\textbf{Introduction}

Quantum magnetism is an active research field for which materials with very large degeneracy of the ground states are sought. Alongside some iridate materials, $\alpha$-RuCl$_3$ \cite{plumb2014alpha, koitzsch2016j} and (more recently) RuBr$_3$ \cite{imai2022zigzag} have been substantially investigated as a leading candidate for physical realisation of the Kitaev model \cite{kitaev2006anyons, takagi2019concept}, albeit that their zigzag magnetic order at low temperatures \cite{Johnson2015Monoclinic} excludes an exact realisation of a quantum spin liquid ground state. Very recently the “cousin” material $\alpha$-RuI$_3$ has been synthesised \cite{ni2022honeycomb, nawa2021strongly}. These experimental reports immediately generated some controversy, even over the basic classification of RuI$_3$ being metallic or insulating, as the interpretation of the resistivity measurements performed on compressed pellets  \cite{ni2022honeycomb, nawa2021strongly} is debated \cite{kaib2022electronic}. It is important to understand the low-energy electronic structure as it is directly related to the achievable magnetic states.

\begin{figure}
\includegraphics{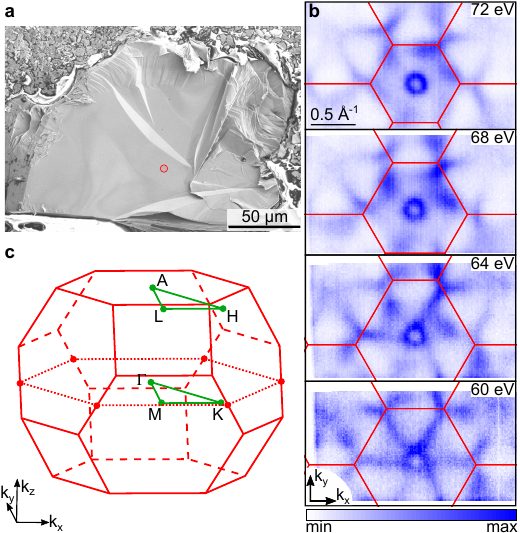}
\caption{Image of the sample and associated ARPES spectrum at different photon energy. \textbf{a} Scanning electron microscopy image of the $\alpha$-RuI$_3$ single crystal after cleaving, showing laminations of the sample. The red circle marks the area measured with the 4 $\mu$m beam spot. \textbf{b} Fermi surfaces measured at selected photon energies corresponding to $k_\text{z}$ between 0 and 3$\pi$/c. The in-plane Brillouin zone boundaries at constant $k_\text{z}$ are overlaid in red. \textbf{c} The 3D Brillouin Zone in red, and the $\Gamma$MK$\Gamma$ and ALHA paths shown in green along which the experimental dispersion has been extracted in Fig.~\ref{Fig:bandDisp_DOS}\textbf{a}.}
\label{Fig:SEM_3DBZ_hvDep}
\end{figure}

Motivated by the experimental ambiguity regarding its ground state, in this paper we investigate the electronic structure of single crystals of honeycomb RuI$_3$ using micro-angle-resolved photoemission spectroscopy (micro-ARPES). First and foremost, we find that RuI$_3$ is, without question, a metal. However we find some fingerprints of electronic correlations in the spectral function, including a modest band renormalisation. Further, the low-symmetry  R$\bar{3}$ space group has a profound impact on the measured electronic structure: despite the crystal structure being formally centrosymmetric and thus achiral, the theoretically calculated and experimentally measured bulk Fermi surfaces, in the $k_\text{x}-k_\text{y}$ plane at any given $k_\text{z}$, appear to be chiral. We describe this property as pseudochiral, in this case associated with the lack of any in-plane mirrors, which in turn is a consequence of the ABC stacking of the honeycomb lattice. Pseudochirality is a concept more generally applicable to ARPES measurements, and we describe the necessary conditions for observing a pseudochiral band structure in ARPES.

\textbf{Results}

\textbf{3D electronic structure of $\alpha$-RuI$_3$}. The small size of the plate-like single crystal samples, grown under pressure, necessitated a few variations to the usual cleave-and-measure ARPES protocols. A fine tungsten wire was used for cleaving the sample (see Methods). Capillary mirror micro-ARPES was used to identify a uniform area with optimal spectral quality; later scanning electron microscopy characterisation (Fig.~\ref{Fig:SEM_3DBZ_hvDep}\textbf{a}) confirmed this to be a flat and homogenous region of the surface. The Fermi Surfaces were obtained by rotating the electron analyser around the fixed sample position, maintaining a constant beam spot on the sample. 

In ARPES, the in-plane momentum ($k_\text{x}$ and $k_\text{y}$) and energy of the initial state are conserved quantities and precisely measurable. However, the out-of-plane momentum ($k_\text{z}$) is not conserved. Nevertheless, within the free electron final state approximation, different planes in $k_\text{z}$ can be probed by varying the energy of the incoming photon \cite{damascelli2004probing}, as shown in Fig.~\ref{Fig:SEM_3DBZ_hvDep}\textbf{b}. We find that 59 and 72~eV correspond to $\Gamma$ and A planes respectively. At 60~eV, very close to a $\Gamma$ plane, we observe a small central pocket connected to six spokes that disperse towards K points, with approximately sixfold rotational symmetry. At 64~eV only three spokes are well-defined, and the symmetry is only threefold. At 68~eV we observe the six-spoke structure in alternating second Brillouin zones, best understood by considering the stacking of the bulk 3D Brillouin zones (Fig.~\ref{Fig:SEM_3DBZ_hvDep}\textbf{c}). At 72~eV, we find the central pocket is the sharpest and brightest feature. The observation of a Fermi surface in all cases is clear evidence that RuI$_3$ is metallic, while the strong variation with photon energy indicates a complex multi-band 3D Fermi surface topology. 

The origin of the three-dimensionality of the Fermi surface can be traced to significant inter-layer hopping, despite the quasi-2D crystal structure. The shortest interlayer bond distance $d_{I-I}=$3.9857~\AA{} is larger than the $d_{Cl-Cl}=$3.6649~\AA{} in RuCl$_3$, but on the other hand, the volume of the I 5p-orbitals are larger than the Cl 3p-orbitals. Thus, the ratio of the interlayer bond distance to the Shannon ionic radius is smaller for RuI$_3$ (1.81) than RuCl$_3$ (2.02), implying significant overlap in RuI$_3$. Therefore we qualitatively expect a stronger interlayer hybridisation in $\alpha$-RuI$_3$ - as is reflected by the $k_\text{z}$-dependence of the ARPES measurements. 

\begin{figure}
\includegraphics{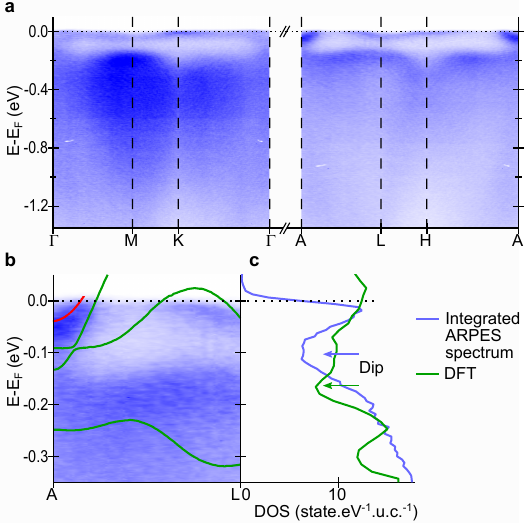}
\caption{Experimental band dispersion and comparison with DFT calculations. \textbf{a} ARPES measurements along the $\Gamma$MK$\Gamma$ and ALHA paths presented on Fig.~\ref{Fig:SEM_3DBZ_hvDep}\textbf{c}. \textbf{b} A zoomed band dispersion on the AL cut shows the presence of a small electron pocket around A (red line) which is predicted by GGA-DFT calculations (green lines). A more comprehensive comparison is available in Supplementary Note A. \textbf{c} Comparison of the DFT-calculated Density of States and the integrated ARPES spectrum (a summation of $\Gamma$M and $\Gamma$K orientation at 72 and 83~eV with linear horizontal and vertical light polarisation, 8 spectra in total).}
\label{Fig:bandDisp_DOS}
\end{figure}

\textbf{Metallicity of $\alpha$-RuI$_3$}. Fig.~\ref{Fig:bandDisp_DOS}\textbf{a} presents the experimental dispersion along the paths $\Gamma$MK$\Gamma$ (h$\nu$=59~eV i.e. $k_\text{z}$=0) and ALHA (h$\nu$=72~eV i.e. $k_\text{z}$=3$\pi$/c), as defined in Fig.~\ref{Fig:SEM_3DBZ_hvDep}\textbf{c}. There are multiple Fermi crossings that derive from two bands as shown in Fig.~\ref{Fig:bandDisp_DOS}\textbf{b}. These two bands have a particularly narrow bandwidth, not dispersing below -0.1~eV anywhere in the Brillouin zone. Experimentally, we find a dip in the integrated spectral weight centered at -0.1~eV. A similar dip is found in the Density of States as calculated by density-functional theory (DFT) at -0.17~eV as shown in Fig.~\ref{Fig:bandDisp_DOS}\textbf{c}, thus we ascribe the experimental dip as mainly a band structure effect.  

It is clearly apparent, however, that the fully occupied bands below the dip are highly broadened compared with the Fermi-crossing bands. In fact, DFT calculations predict numerous valence bands and continuous density of states \cite{zhang2022theoretical} for several eV, but we find that below $\sim$-0.3~eV one could not resolve any distinct dispersions within a very broad spectrum. 

One contribution to the observed broadening of ARPES spectra is the disorder induced by Ru on the vacancy site, and deficiency of Ru atoms on the metal site, as reported in the sample characterisation \cite{ni2022honeycomb}. However the contribution of disorder to spectral broadening is generally expected to be energy independent, and while it may be limiting the sharpness of features at $E_\text{F}$, it cannot explain the energy dependence of the broadening. The $k_\text{z}$ broadening of the 3D-like band structure also plays a role, but by itself cannot explain the absence of any resolvable features at all. The energy scale of the observed broadening of ARPES spectra is also not compatible with electron-phonon coupling. 

\begin{figure*}
\includegraphics{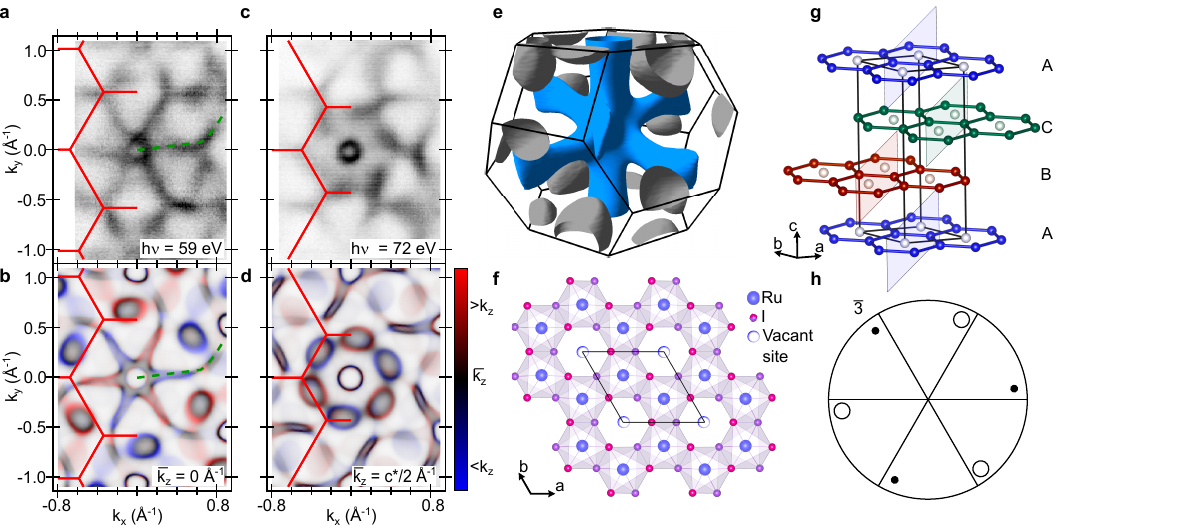}
\caption{Experimental Fermi surfaces compared with DFT and explanation of the pseudochirality origin though the layer stacking and space group. \textbf{a}, \textbf{c} Fermi surface measured at 59 and 72~eV, corresponding to $k_\text{z}$ = 0 and 3$\pi$/c, respectively. \textbf{b}, \textbf{d} Fermi surface calculated by DFT at $\bar{k_\text{z}}$=0 and 3$\pi$/c, respectively. A Lorentzian energy broadening (2$\gamma_E$=30~meV) and Lorentzian $k_\text{z}$ broadening (2$\gamma_k$=1.2~nm$^{-1}$) has been added to reproduce realistic experimental resolution. Red and blue color indicate if the intensity comes mainly from a band above or below $\bar{k_\text{z}}$. The dashed green line in \textbf{a} and \textbf{b} is a guide to the eye, identical on both panels. \textbf{e} 3D representation of the Fermi surface in the Brillouin Zone. In blue is the electron pocket and in grey are the holes pockets. \textbf{f} Single layer of $\alpha$-RuI$_3$. The pink and purple iodine atoms represent the iodine above and below the Ru-plane, respectively. The black line is the unit cell in the ab-plane. The nominally vacant sites have 2\% Ru occupancy\cite{ni2022honeycomb}. \textbf{g} Stacking along the c-axis of Ru hexagonal-lattice. The almost white atoms represent the nominally vacant sites' locations similar to \textbf{f}. The planes along (1-10) direction are almost symmetry planes of each individual layer, but due to the stacking, the crystal doesn't have any global mirror (or glide) symmetry. \textbf{h} Representation of the $\bar{3}$ Laue class' equivalents points.}
\label{Fig:Struct_Sym_FS}
\end{figure*}

Instead, the energy dependency suggests that local electronic correlations contribute significantly to the spectral broadening. The situation is qualitatively similar to other correlated Ru-compound such as Sr$_2$RuO$_4$ \cite{tamai2019high} or Ca$_3$Ru$_2$O$_7$ \cite{markovic2020electronically}; especially in the latter, one also observes a few sharp bands close to $E_\text{F}$, giving way to a blur of continuous spectral weight below $\sim$-0.1~eV. Thus the broadening gives the first fingerprint of local electronic correlations.  

Fig.~\ref{Fig:bandDisp_DOS}\textbf{b} presents a zoom along the AL path, where the small electron pocket centred around A is sharp enough to perform quantitative analysis. The sharpness of this band, relative to the other branches near $E_\text{F}$, comes from its minimal $k_\text{z}$ dispersion around $k_\text{z}$=3$\pi$/c, as visible in Fig.~\ref{Fig:Struct_Sym_FS}\textbf{e}. The experimental Fermi wavevector $k_\text{F}$ is 0.075~\AA$^{-1}$ while the one calculated by GGA-DFT is 0.120~\AA$^{-1}$, possibly due to a net hole doping from the Ru deficiency, or more simply an inaccuracy of the DFT calculation. The measured Fermi velocity is 0.94~eV$\cdot$\AA{} whereas the DFT velocity at the same k-vector is 1.42~eV$\cdot$\AA{} which corresponds to a band renormalisation factor of $\eta$=1.52. For comparison, Sr$_2$RuO$_4$ has a band renormalisation between 3 and 5 depending on the band and position \cite{tamai2019high}. Thus we conclude that RuI$_3$ displays signatures consistent with a moderate degree of electronic correlations. 

\textbf{The pseudochiral Fermi surface.}
The measured Fermi surface in the $k_\text{z}$=0 plane shown in Fig.~\ref{Fig:Struct_Sym_FS}\textbf{a} is sixfold symmetric, but the spokes emanating from the $\Gamma$ point towards K points are not precisely mirror symmetric, and do not intersect the Brillouin zone boundary exactly at K. The mirror symmetry breaking in the Fermi surface is accentuated by second derivative and curvature analysis \cite{zhang2011precise}, available in Supplementary Note E. The mirror symmetry-breaking cannot be attributed to the measurement geometry (see methods), since the set-up preserves a mirror plane which maps $k_\text{y}$ to $-k_\text{y}$, which is nevertheless broken in the data. 

To elucidate this observation, we simulate the ARPES spectrum using the DFT calculation by integrating 2D ($k_\text{x},k_\text{y}$) slices over a Lorentzian distribution in $k_\text{z}$ (FWHM=1.2~nm$^{-1}$) and in energy (FWHM=30~meV), accounting for the intrinsic $k_\text{z}$ broadening and the experimental energy resolution respectively. The result of the convolved DFT calculation is visible in Fig.~\ref{Fig:Struct_Sym_FS}\textbf{b} and \textbf{d} for $k_\text{z}$=0 and $k_\text{z}$=3$\pi$/c, respectively. The color scale describes the low (white) / high (black) intensity areas whereas the red/blue colors indicates the $k_\text{z}$ origin of the contributions (red $k_\text{z}>\bar{k_\text{z}}$, blue $k_\text{z}<\bar{k_\text{z}}$). The main features of the measurements in both the $k_\text{z}$=0 and 3$\pi$/c planes are reproduced, including their symmetry as visible in Fig.~\ref{Fig:Struct_Sym_FS}(\textbf{a}-\textbf{d}). In the $\Gamma$ and A planes, the Fermi surface is sixfold symmetric, but for a given $k_\text{z}$ plane only threefold rotation is expected, as seen in the 3D Fermi surface in Fig.~\ref{Fig:Struct_Sym_FS}\textbf{e}. Importantly, both the simulations and the experiment show rotational symmetry, but no mirror planes. 

The absence of mirror planes can be understood from a real space perspective. A monolayer of an idealised honeycomb lattice of RuI$_3$ could host 3 mirror planes, passing through the voids of the honeycombs (see Fig.~\ref{Fig:Struct_Sym_FS}\textbf{f}). However, if we consider ABC stacking of the honeycomb lattice, where each layer is shifted by ($\frac{-1}{3},\frac{1}{3},\frac{1}{3}$) with respect to the layer beneath, the void positions do not align between layers, thus no global mirror planes can exist, as illustrated in Fig.~\ref{Fig:Struct_Sym_FS}\textbf{g}. Therefore, the combination of the honeycomb lattice (in its physical realisation with cations alternatively above and below the plane) with ABC stacking excludes the existence of any mirror planes. RuI$_3$ thus forms in space group R$\bar{3}$ (No 148), where the symmetry elements correspond only to the $\bar{3}$ rotoinversion operator (see Fig.~\ref{Fig:Struct_Sym_FS}\textbf{h}). It is formally achiral as it is centrosymmetric, but has no mirror or glide-mirror symmetries. 

In 3D real space, a crystal is to be considered chiral if its space group contains only proper operations\cite{fecher2022chirality}, i.e. does not contain inversion, mirror, glide, or rotoinversion.
The conditions to obtain true chirality in 3D k-space are even more stringent since even if inversion symmetry is not present in the crystal structure, it is restored in k-space by time-reversal symmetry.
RuI$_3$ has both inversion symmetry in real space and does not break time reversal. Thus RuI$_3$ is achiral in the crystallographic definition, and we also cannot describe the electronic structure of RuI$_3$ as chiral in a 3D sense. 

Nevertheless, both the simulated and experimental Fermi surfaces projected in the 2D $k_\text{x},k_\text{y}$ plane lack any mirror symmetry. This motivates us to introduce the concept of \textit{pseudochirality}: the case that a 2D projection (relative to some given surface direction) of a bulk 3D band structure does not contain any mirror symmetry. 
 
More generally, we can ask the question: under what conditions can pseudochirality be observed in a bulk-sensitive ARPES measurement? Formally, the symmetry of electronic band dispersions in $k$-space is determined by the Laue class of the material. The Laue classes are eleven geometric crystal classes containing centrosymmetric crystallographic types of point groups and their subgroups\cite{shmueli2016methods, vacher2003brillouin}. Of the 11 Laue classes, $\bar{3}$ (the case for RuI$_3$, shown in Fig.~\ref{Fig:Struct_Sym_FS}\textbf{h}) and $\bar{1}$ have no mirrors at all, thus materials which form in space groups with these Laue classes should generically show pseudochiral electronic structure, independent of the cleavage plane. 

For the cases $6/m$, $4/m$ and $2/m$ where the mirror plane is perpendicular to the rotation axis, if we assume a cleave plane perpendicular to the rotation axis, there would be no mirrors within a 2D projection in the $k_\text{x}$ - $k_\text{y}$ plane, and again the electronic structure will be pseudochiral. In fact, pseudochirality can, theoretically, be observed in any material; one simply has to choose a termination such that the surface normal does not lie in any mirror planes of the Laue class. There is a close relationship between the conditions for pseudochirality and surface chirality\cite{pratt2005symmetry, shukla2020chiral}, although they are not exactly equivalent, as discussed further in Supplementary Note C.

Despite the many theoretical routes to pseudochirality as discussed further in Supplementary Note D, RuI$_3$ has a few characteristics which, practically speaking, assist with the observation of pseudochiral electronic structure. First, single crystals of good quality exist, can be cleaved, and are stable in UHV. Second, it is metallic; having sharp quasiparticle bands at $E_\text{F}$ will generally help with observing subtle deviations from the symmetric case. Third, having large Fermi surfaces also helps, as the deviations from mirror conditions become more apparent away from the $\Gamma$ point. In principle, the top/bottom faces of the same crystal of RuI$_3$ would present mirror-image spectra in 2D ARPES measurements (i.e. appearing left/right handed), in otherwise equivalent conditions. 

To our knowledge, there is only one previously reported system with pseudochirality in measurements of bulk electronic structure, which is the $\sqrt{13}\times{}\sqrt{13}R13.9^\circ$ phase of 1T-TaS$_2$ and TaSe$_2$ \cite{yang2022visualization, bovet2004pseudogapped, ngankeu2017quasi}. This is a special case where the mirror symmetry of the high-temperature $P\bar{3}m1$ (No 164) phase is broken by the star-of-David superstructure; the band structure is reconstructed and the ground state also has Laue class $\bar{3}$, allowing pseudochiral electronic structure. 

A conceptual difference between 1T-TaS$_2$ and RuI$_3$ is that in the former case, there is a strong breaking of in-plane mirrors by the in-plane superstructure, while in the case of RuI$_3$, it is rather the out-of-plane stacking that breaks the mirror symmetry. An alternative polytype of RuI$_3$, grown under different conditions, has been reported in space group P$\bar{3}$1c (No 163) \cite{nawa2021strongly}, differing principally in having an alternating two-layer stacking configuration. Due to the additional $2/c$ glide-mirror symmetry element, the Laue class is $\bar{3}$1m, which includes a mirror. Consistent with this, the calculated Fermi surface for P$\bar{3}$1c phase is qualitatively similar to the $R\bar{3}$ phase studied here, but not pseudochiral, having a mirror plane parallel to $\Gamma-M$ (see Supplementary Note B). Thus in the context of honeycomb lattice systems, it is the ABC stacking that specifically enables pseudochiral electronic structure (for $\{$001$\}$ cleavage, which is the only realistic possibility). 

\textbf{Discussion}

The first outcome of this work is confirming the metallicity of $\alpha$-RuI$_3$. The multi-bands 3D Fermi surface revealed by ARPES in Fig.~\ref{Fig:SEM_3DBZ_hvDep}\textbf{b} clearly support the metallic character of RuI$_3$ reported by transport measurements \cite{ni2022honeycomb, nawa2021strongly} on compressed pellets or polycrystalline samples. Due to the 3D nature of the Fermi surface, if single crystal resistivity measurements were performed, we expect the resistivity anisotropy ratio between in-plane and out-of-plane to be close to unity, unlike the three orders of magnitude reported in RuCl$_3$ \cite{binotto1971optical}. 

While some theoretical work predicted $\alpha$-RuI$_3$ to be possibly insulating, other studies found that only an unrealistically large U repulsion could open a Mott gap \cite{zhang2022theoretical}, instead expecting a metallic ground state. Our results support the latter scenario, while also finding some spectral signatures of electronic interactions through the energy-dependent broadening and band renormalisations. Clearly, the on-site interactions, although suppressed compared with the Mott insulating $\alpha$-RuCl$_3$ due to the increased hybridisation with the I orbitals, are not completely quenched in $\alpha$-RuI$_3$. It would be interesting to see if Dynamical mean-field theory calculations can shed further light on the moderately correlated state we find experimentally. 

Although we have shown that there are many routes in theory, identifying practical realisations of pseudochirality is challenging. We suggest four avenues that could be fruitful: (1) other materials in space group 148, $R\bar{3}$, which is fairly common in nature and where a layered structure with good cleavage properties can be expected; (2) looking for materials with Laue class $4/m$ and $6/m$ which might have (001) cleavage; (3) The $\{$111$\}$ surfaces of cubic systems with Laue class $m\bar{3}$; (4) Deliberate miscuts of elemental metals, such as (321) or (531), using soft x-ray ARPES to access the bulk rather than the surface states.

The concept of pseudochirality is introduced here primarily for the photoemission spectroscopy community, providing a framework for understanding when bulk-sensitive ARPES measurements can appear chiral. However, other physical responses and techniques are likely to be dependent on the established concepts of real-space crystallographic chirality, or surface chirality. 

\textbf{Conclusions}

In conclusion, ARPES measurement on single crystal of $\alpha$-RuI$_3$ reveals a metallic 3D band structure and signatures of moderate electronic correlations. The Fermi surface is close to the one calculated by DFT with a complex-shaped electron and a hole-like pockets. As in the calculations, the experimental 2D Fermi surface measured at any $k_\text{z}$ exhibits at least threefold rotational symmetry, but never a mirror plane. We thus propose $\alpha$-RuI$_3$ as an exemplar material for electronic pseudochirality.  

\textbf{Methods}

High-quality $\alpha$-RuI$_3$ crystals were synthesised by heating amorphous RuI$_3$ powder at high pressure and subsequently purified by vapor transport techniques in a quartz tube. The crystallographic phase was characterised by single-crystal and powder X-ray diffraction at room temperature confirming the three-layer centrosymmetric rhombohedral symmetry structure, with a honeycomb-layer lattice. The structural refinement indicates a slightly Ru-deficient composition Ru$_{0.98}$I$_3$ with 96\% occupation of the Ru sites and 2\% occupation of the ideal vacant site\cite{ni2022honeycomb}. To optimise the surface preparation procedure, a few crystals were mounted on the same sample plate with a 50~$\mu$m W wire glued on the top to cleave the surfaces in UHV. The typical size of the sample was 200~$\mu$m as visible on Fig.~\ref{Fig:SEM_3DBZ_hvDep}\textbf{a}. The ARPES measurements were performed at the nano-ARPES branch of the Diamond I05 beamline \cite{hoesch2017facility} using a DA30 analyser with vertical slit corresponding to $k_\text{y}$ in the experimental data. All the constant energy cuts displayed have been measured with linear horizontal light polarisation. A capillary mirror has been used to focus the photon beam down to 4~$\mu$m. No sign of degradation was observed after 48 hours in 10$^{-10}$~mbar pressure at 40~K temperature.

Density function theory (DFT) calculations have been performed using generalized gradient approximations (GGA) functional with spin-orbit coupling (SOC) but without Coulomb repulsion U. The calculations have been done with WIEN2K \cite{blaha2020wien2k} using the room temperature structure \cite{ni2022honeycomb}.

\textbf{Data availability}

The authors declare that the main data supporting the findings of this study are available within the paper and its Supplementary Information. Extra data are available from the corresponding authors upon request.

\bibliography{main}

\begin{acknowledgments}
\textbf{Acknowledgments}

 Niels B. M. Schröter is warmly thanked for his constructive comments as well as Anna Warren for the access to scanning electron microscopy imaging. We acknowledge Diamond Light Source for time on Beamline i05-1 under Proposal NT31246. We are grateful for the Computing resources provided by STFC Scientific Computing Department’s SCARF cluster. The crystal synthesis was founded by the Gordon and Betty Moore Foundation, EPiQS initiative, Grant No. GBMF-9066.
\end{acknowledgments}

\textbf{Competing interests}\\
The authors declare no competing interests.

\textbf{Author Contributions}\\
DN and RJC synthesised and characterised the samples. AL, MDW, TKK and CC performed experimental measurements on the samples and contributed to the analysis of the experimental data and manuscript writing.

\end{document}